\documentclass[%
preprint,
 amsmath,amssymb,
 aps,
]{revtex4-1}

\usepackage{bm}
\usepackage{graphicx}
\usepackage{subfigure}
\usepackage{color}

\begin{document}

\title{Nucleation of second-phase near elastic defects in crystalline solids}

\author{C. Bjerk\'en and A. R. Massih}
 \altaffiliation[Also at ]{Quantum Technologies AB, Uppsala Science Park, SE-751 83 Uppsala, Sweden}
\affiliation{%
Division of Materials Science, School of Technology, Malm\"{o} University, SE-205 06 Malm\"{o}, Sweden\\
}%




\date{\today}

\begin{abstract}

The problem of heterogeneous nucleation of second-phase in alloys in the vicinity of elastic defects is considered. The defect can be a dislocation line or a crack tip residing in a crystalline solid. We use the Ginzburg-Landau equation to describe the spatiotemporal evolution of the order parameter in the environs of the defect. The model accounts for the elasticity of the solid and the interaction of order parameter field with the elastic field of the defect. A finite volume numerical method is used to solve the governing partial differential equation for the order parameter. We examine the nature of the phase transition in the vicinity of the defects.

\end{abstract}

\maketitle

\section{Introduction}

The presence of elastic defects such as dislocations and cracks may induce nucleation of a second phase in many alloys \cite{Boulbitch_Toledano_1998,Leonard_Desai_1998}. For example, the formation of brittle hydrides in titanium and zirconium alloys (TiH$_x$, ZrH$_x$) is of special interest for aerospace and nuclear industries, since they may cause embrittlement of these alloys used in various equipment \cite{Coleman_2003}. Hydride formation is commonly accompanied by a preferred orientation of the precipitates (platelets) due to the texture of the polycrystalline material and/or the presence of external stress \cite{Massih_Jernkvist_2009}. Moreover, the crystal structure of the hydride (face-centered cubic for $\delta$-hydride) differs from that of the matrix (hexagonal close-packed for  $\alpha$-Ti). In the present study, we use a phase field approach to analyze the nature of the structural part of the new phase formation near elastic defects. That is, the effect of composition is not included in our analysis. The model used here rests on the Ginzburg-Landau theory of phase transition in which a scalar non-conserved order parameter characterizes the presence or the absence of the nucleus. The interaction between the order parameter and the deformation field is also taken into account \cite{Larkin_Pitkin_1969,Imry_1974}. A more general set-up was presented in \cite{Massih_2011a}.

\section{Model description}
\label{sec:gov-eqs}

We consider two types of elastic defects in a crystalline material; an edge dislocation and a semi-infinite crack. The considered phase transformation is the nucleation of a second phase in an elastic material. A single structural order parameter   that accounts for the symmetry of structure is used to characterize the phases. It is supposed to be a scalar field (Ising model)   $\eta(\mathbf{r},t)$ that is a function of space $\mathbf{r}$ and time $t$. Hence,  $\eta=0$ corresponds to solid solution and $\eta \neq 0$ to a nucleus. The total free energy of the system is written \cite{Massih_2011a}
\begin{eqnarray}
\mathcal{F} = \mathcal{F}_{st}+\mathcal{F}_{el}+\mathcal{F}_{int},
\label{eqn:fe-tot}
\end{eqnarray}
\noindent
where $\mathcal{F}_{st}$ is the structural free energy, $\mathcal{F}_{el}$ the elastic strain energy, and  $\mathcal{F}_{int}$ is the interaction energy between the structural order parameter and the strain field. The structural free energy is
\begin{eqnarray}
\mathcal{F}_{st}= \int
\big[\frac{g}{2}(\nabla\eta)^2+\mathcal{V}(\eta)\big]d\mathbf{r},
\label{eqn:f-str}
\end{eqnarray}
where the space integral is within the volume of the system $d\mathbf{r}=\rho d\rho d\theta dz$. Here  $g(\nabla\eta)^2$ accounts for the spatial dependence of the order parameter, $g$ is a positive constant, and  the second term in the integrand is the Landau potential \cite{Landau_Lifshitz_1980}
\begin{eqnarray}
\mathcal{V}(\eta) = \frac{1}{2}r_0\eta^2 + \frac{1}{4}u_0\eta^4 + \frac{1}{6}v_0\eta^6,
\label{eqn:landau-pot}
\end{eqnarray}
where $r_0$ is taken to be a linear function of temperature $T$, e.g. $r_0=\alpha_0(T-T_c)$, $\alpha_0$ is a positive constant and $T_c$ the transition temperature in the absence of elastic coupling. The coefficients $u_0$ and $v_0$ are considered to be temperature independent. The elastic free energy is
\begin{eqnarray}
\mathcal{F}_{el}= \int \Big[\frac{K}{2}\big(\nabla \cdot \mathbf{u}\big)^2
+ M\sum_{ij}\Big(u_{ij} - \frac{\delta_{ij}}{d}\nabla \cdot \mathbf{u}\Big)^2\Big]d\mathbf{r},
\label{eqn:f-el}
\end{eqnarray}
\noindent
where $K$ and $M$ are the bulk and shear modulus, respectively, $u_{ij}=(\nabla_j u_i+\nabla_i u_j)/2$ is the strain tensor with $\nabla_i\equiv \partial/\partial x_i$, $d$ the space dimensionality, and $i,j$ stand for $x,y,z$ in $d=3$ ($x,y$ in $d=2$). Finally, the interaction energy is
\begin{equation}
\mathcal{F}_{int} =\kappa\int  \eta^2\, \nabla \cdot \mathbf{u}\, d\mathbf{r},
\label{eqn:f-int}
\end{equation}
where  $\eta^2\nabla \cdot \mathbf{u}$ describes the interaction between the order parameter and the deformation. The strength of this interaction is denoted by $\kappa$  and is taken to be a constant.

\begin{figure}[htbp]
\begin{center}
\subfigure[\hspace{1mm} Edge dislocation]
{
\label{fig:geom-disl}
\includegraphics{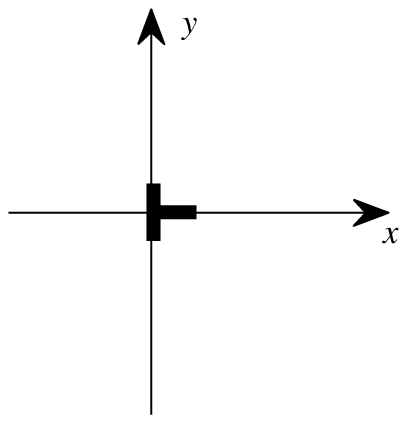}
}
\hspace{1cm}
\subfigure[\hspace{1mm} Semi-infinite crack]
{
\label{fig:geom-crack}
\includegraphics{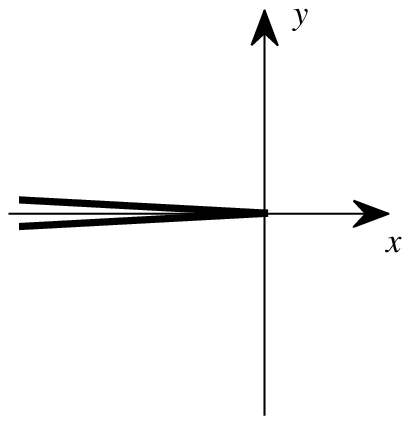}
}
\caption{Geometry of the defects; $x=\rho\cos\theta$  and $y =\rho\sin\theta$.}
\label{fig:geom-defect}
\end{center}
\end{figure}

The temporal evolution of the spatial order parameter is determined by solving the time-dependent Ginzburg-Landau equation for a non-conserved field, cf. \cite{Lifshitz_Pitaevskii_1981},
\begin{equation}
\label{eqn:tdgl-eq}
\frac{\partial \eta}{\partial t}  =  -L_a\frac{\delta\mathcal{F}}{\delta \eta},
\end{equation}
where $L_a$ is a kinetic coefficient that characterizes the interface boundary mobility. Defects in a crystalline material, such as dislocations and cracks, render internal strains which change the equilibrium condition in the solid. If $\mathbf{f}(\mathbf{r})$  denotes the variation of the strain field due to the defect, the equilibrium condition that includes the force field is generated by \cite{Landau_Lifshitz_1970}
\begin{eqnarray}
M\nabla^2 \mathbf{u}+(\Lambda-M)\nabla \nabla \cdot \mathbf{u} + \kappa\nabla \eta^2 = M \mathbf{f}(\mathbf{r}),
\label{eqn:mecheq}
\end{eqnarray}
where $\Lambda=K+2M(1-1/d)$. Equation (\ref{eqn:mecheq}) is then used to eliminate the elastic field from the expression for the total free energy, which now can be expressed as
\begin{equation}
\mathcal{F}[\eta] = \int \Big[\frac{g}{2}(\nabla\eta)^2+\frac{1}{2}r_1\eta^2 + \frac{1}{4}u_1\eta^4 + \frac{1}{6}v_0\eta^6\,\Big] d\mathbf{r}.
\label{eqn:toten1}
\end{equation}
The last three terms in Eq. (\ref{eqn:toten1}) correspond to the Landau potential energy in Eq. (3) but with modified coefficients:
\begin{eqnarray}
r_1 &=& r_0-\kappa A\cos\theta/\rho, \quad \text{for an edge dislocation},\\
\label{eqn:r1-crack}
r_1 &=& r_0-\kappa B\cos(\theta/2)/\rho^{1/2}, \quad \text{for a crack},\\
\label{eqn:r1-disloc}
u_1 &=& u_0-2\kappa^2/\Lambda,
\label{eqn:u1-elastic}
\end{eqnarray}
\noindent
with $A=(2b/\pi)M/\Lambda$ and $B=4K_\mathrm{I}(1-2\nu)(1+\nu)/(2\pi)^{1/2}E$, where $b$ is the magnitude of the Burgers vector, $K_\mathrm{I}$ the mode I stress intensity factor, $\nu$  Poisson's ratio, and $E$ Young's modulus of the material. For a defect free crystal $r_1=r_0$, and for a rigid crystal $u_1=u_0$.  Figure\ref{fig:geom-defect} shows the geometry of the two defects.

\begin{figure}[htbp]
\begin{center}
\subfigure[\hspace{1mm} $u_1 > 0$]
{
\label{fig:LandauUp}
\includegraphics{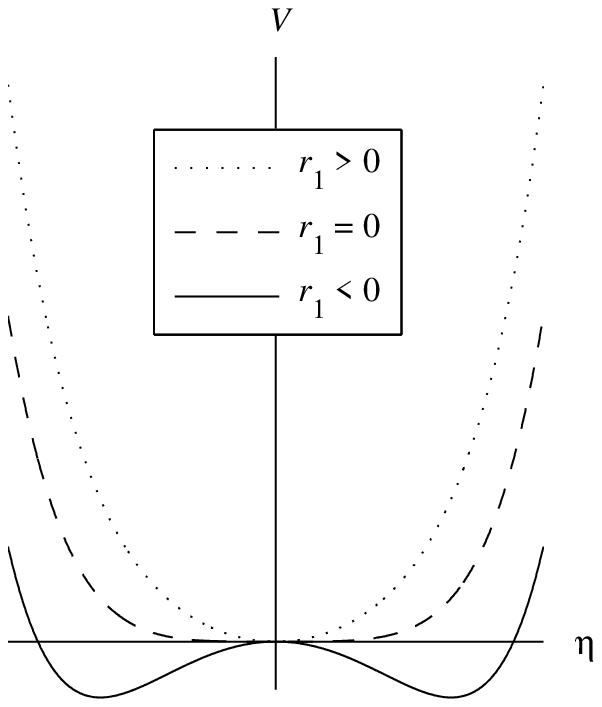}
}
\hspace{1cm}
\subfigure[\hspace{1mm} $u_1 < 0$]
{
\label{fig:LandauUn}
\includegraphics{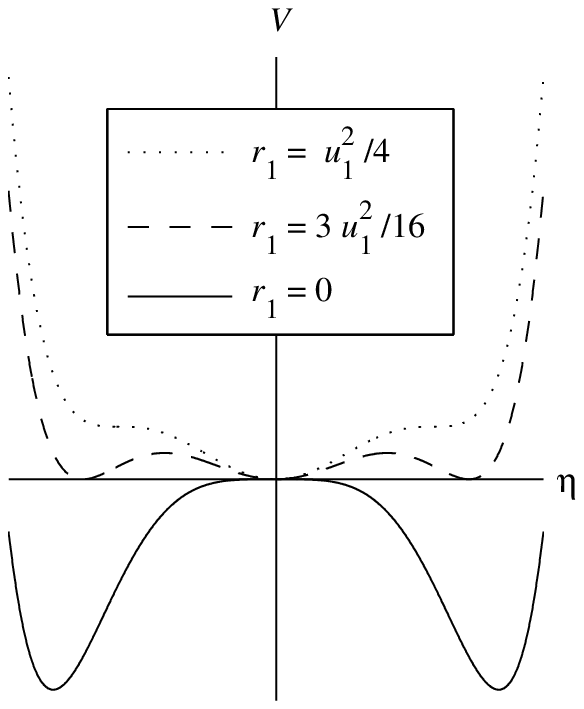}
}
\caption{The Landau potential $\mathcal{V}(\eta) = \frac{1}{2}r_1\eta^2 + \frac{1}{4}u_1\eta^4 + \frac{1}{6}v_0\eta^6$ with $v_0>0$.}
\label{fig:Landau-pot}
\end{center}
\end{figure}

Inserting Eq. (\ref{eqn:toten1}) into Eq. (\ref{eqn:tdgl-eq}), gives us the governing equation for the space-time variation of the order parameter
\begin{equation}
\label{eqn:tdgl-eq2}
\frac{1}{L_a}\frac{\partial \eta}{\partial t}  =  g\nabla^2\eta-\big(r_1\eta + u_1\eta^3+v_0\eta^5\big).
\end{equation}
\noindent
At equilibrium, $\partial\eta /\partial t = 0$, and the Landau potential, i.e. Eq. (\ref{eqn:landau-pot}), tells us the kind of phase transition plus for which set of parameters nucleation and growth will take place. The Landau potential with the modified coefficients $r_1$ and $u_1$, and $v_0 > 0$ for two sets of parameters are shown in Fig. \ref{fig:Landau-pot}, one with $u_1 > 0$ and the other with $u_1 < 0$. If $u_1 > 0$ and $r_1 > 0$, no nucleation of the second phase will occur, however,  $r_1 = 0$ embodies  an onset of nucleation, and for $r_1 < 0$ the nuclei will continue to grow since only the second phase is stable, see Fig. \ref{fig:LandauUp}. In the case of $u_1 < 0$ (Fig. \ref{fig:LandauUn}), a metastable second phase may emerge if $r_1 < u_1^2/4v_0$ since the potential then has two local minima at $\eta \ne 0$. For $r_1 = 3u_1^2/16v_0$, both phases are equally stable and so may coexist. For even a smaller positive value of $r_1$ the solid solution is metastable and the second phase stable. When $r_1 = 0$ only the second phase will exist.

\begin{figure}[htbp]
\begin{center}
\subfigure[\hspace{1mm} $ R_0 > 0$]
{
\label{fig:DislRposUneg}
\includegraphics{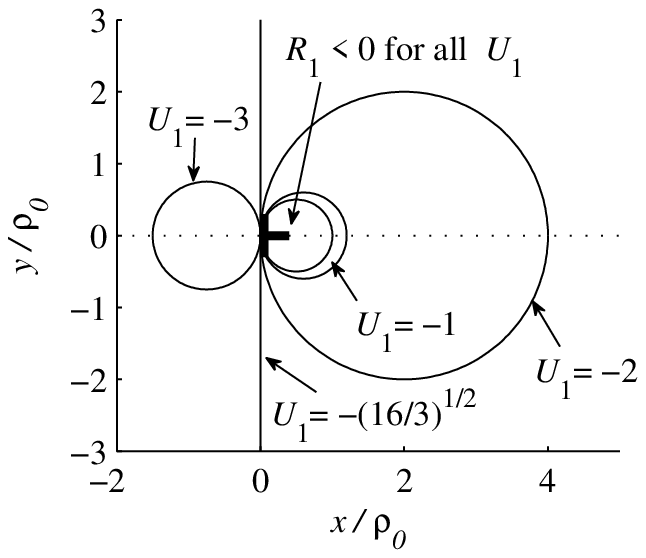}
}
\hspace{1cm}
\subfigure[\hspace{1mm} $R_0 > 0$]
{
\label{fig:DislRnegU}
\includegraphics{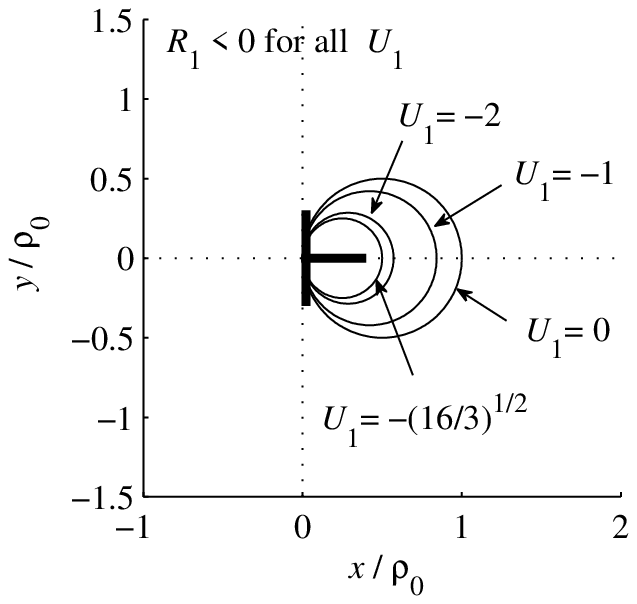}
}
\\
\subfigure[\hspace{1mm} $R_0 > 0$]
{
\label{fig:CrackRposUneg}
\includegraphics{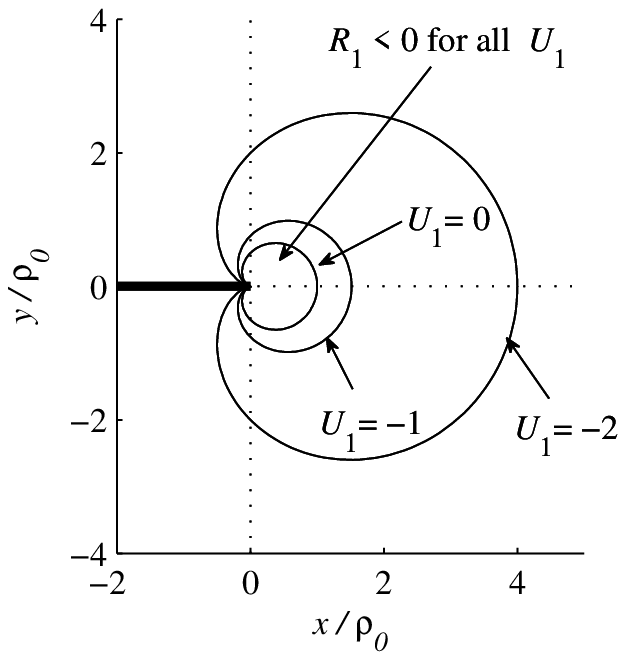}
}
\hspace{1cm}
\subfigure[\hspace{1mm} $R_0 > 0$]
{
\label{fig:CrackRnegU}
\includegraphics{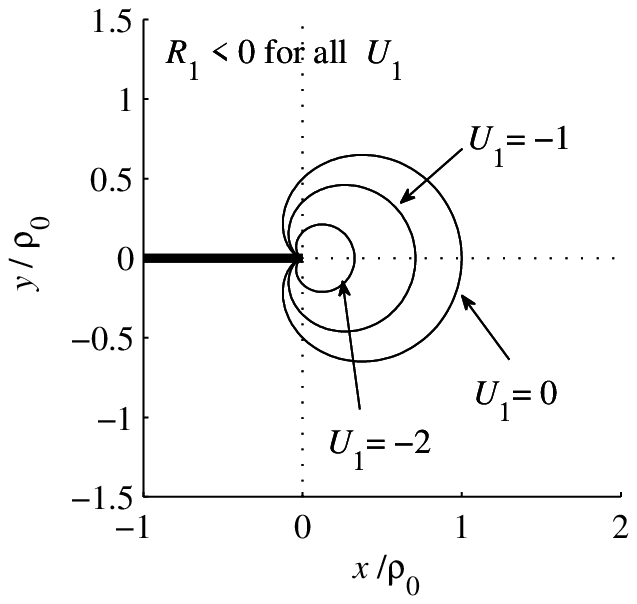}
}
\caption{Characteristic contours in the presence of a dislocation (a)-(b), and a crack (c)-(d); where $R_1 =3U_1^2/16$ for $U_1\leq0$.}
\label{fig:Defect-character}
\end{center}
\end{figure}

Since $r_1$ is a function of both temperature and spatial coordinates, the location where nucleation and further evolution of the second phase will occur, relative to that of the defect, will vary with the choice of $u_1$. In Fig. \ref{fig:Defect-character}, the boundaries corresponding to triple minima, i.e. $r_1 = 3u_1^2/16$ when $u_1\leq 0$, for the dislocation and the semi-infinite crack are shown, respectively. In the figure, normalized coefficients $U_1$ and $R_1$ corresponding to $u_1$ and $r_1$ are used, which are defined below. In Figs. \ref{fig:DislRposUneg} and \ref{fig:CrackRposUneg} with $R_0 > 0$, there are areas close to defect where $R_1 < 0$, and thus a second phase is expected to nucleate there for all values of $U_1$.

\section{Numerical method}
\label{sec:numeric}
For the numerical simulations, we have used the open-source program FiPy \cite{Guyer_et_al_2009}  which is an object oriented, partial differential equation solver based on a standard finite volume approach. A square 2D-mesh consisting of $200\times200$ equally sized elements is used for the computations, where each element has a side length $\Delta l = \rho_0/20$, with  $\rho_0$ being a characteristic length that is introduced below. The simulations are performed using a constant time increment  $\Delta t = 0.9\Delta l^2/(2g)$. Periodic boundary conditions are applied, and the initial value of the order parameter is taken to be a small random number between 0.005 and 0.01. The dislocation is located at the origin, $(x,y) = (0,0)$, with the slip direction along the line $x = 0$, see Fig. \ref{fig:geom-disl}. The crack lies along the negative part of the $x$-axis with the tip at the origin, see Fig. \ref{fig:geom-crack}. The system is considered to be large enough to model a single dislocation and a long single crack, respectively, without being affected by the periodic boundary conditions. For the computations, Eq. (\ref{eqn:tdgl-eq2}) is rewritten as
\begin{equation}
\label{eqn:tdgl-reduced}
\frac{\partial \eta}{\partial t}  =  \nabla^2\eta-\big(R_1\eta + U_1\eta^3+\eta^5\big),
\end{equation}
\noindent
where $t \equiv L_a v_0t$, $x \equiv (v_0/g)^{1/2}x$ and $y \equiv (v_0/g)^{1/2}y$. The coefficients are now $R_1 = R_0 (1-\rho_0 \cos\theta/\rho)$ for the dislocation with $R_0 = r_0/v_0$ and  $\rho_0 = \kappa A/r_0$; whereas $R_1 = R_0[1-\cos(\theta /2)/(\rho/\rho_0)^{1/2}]$,  $\rho_0 = (\kappa B/r_0)^2$ for the crack, and $U_1 = u_1/v_0$. Thus a characteristic length,  $\rho_0$, is introduced. In this parametric study, $R_0$ and $U_1$ are varied and  $\rho_0$ is set equal to unity. The time steps are taken to be small enough to capture the evolution for all the different combinations of $R_0$ and $U_1$ that are studied. $R_0$ is a linear function of the temperature difference, as defined above, and thus quenching the system to below the transition temperature means $R_0 < 0$; and for $T > T_c$, $R_0 > 0$.

\section{Results and Discussion}
\label{sec:results}
Results from simulations of the evolution of in the vicinity of a dislocation are first given for some combinations of $R_0$ and $U_1$. Thereafter, the outcome from the calculations with a crack in the system is alluded. The details for the latter case will be presented elsewhere.
\subsection{Dislocation: $R_0 =-1$ and $U_1=0$}
\begin{figure}[htbp]
\begin{center}
\subfigure[\hspace{1mm} $ t=100\Delta t $]
{
\label{fig:Surf-Time100}
\includegraphics{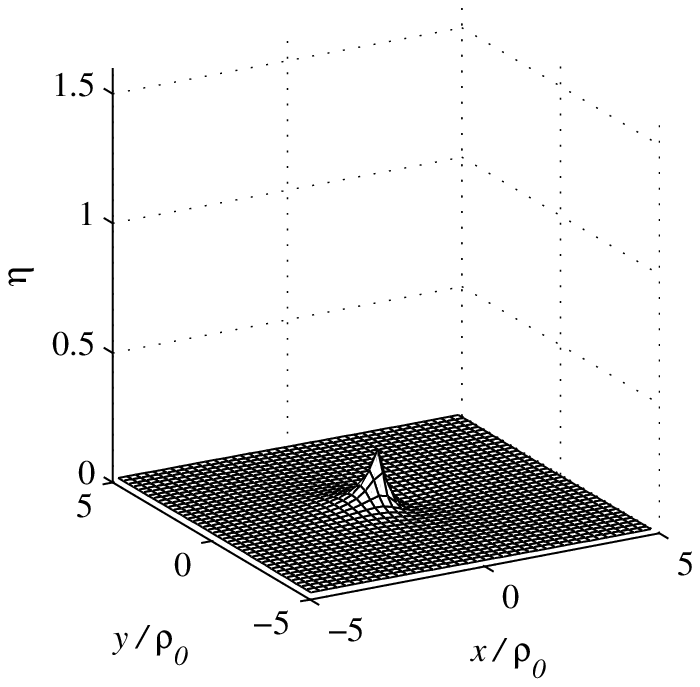}
}
\hspace{1cm}
\subfigure[\hspace{1mm} $ t=150\Delta t $]
{
\label{fig:Surf-Time150}
\includegraphics{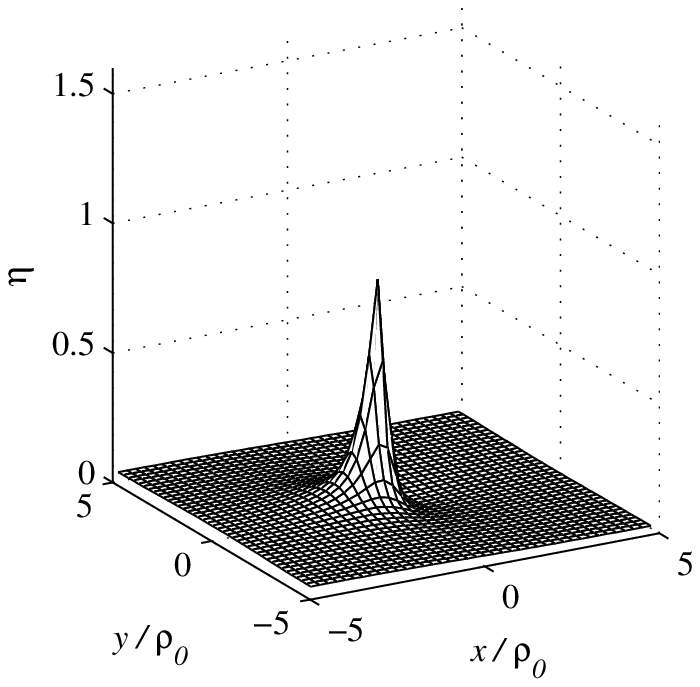}
}
\\
\subfigure[\hspace{1mm} $ t = 200\Delta t $]
{
\label{fig:Surf-Time200}
\includegraphics{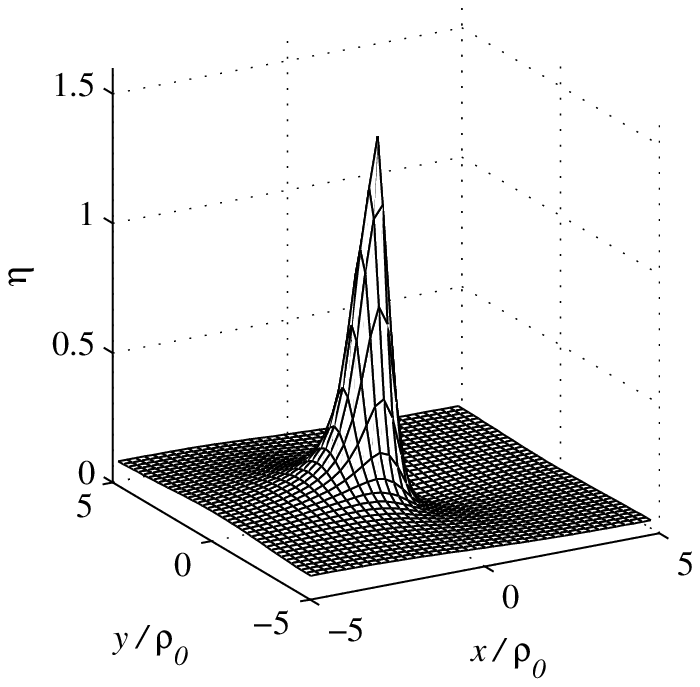}
}
\hspace{1cm}
\subfigure[\hspace{1mm} $ t=250\Delta t $]
{
\label{fig:Surf-Time250}
\includegraphics{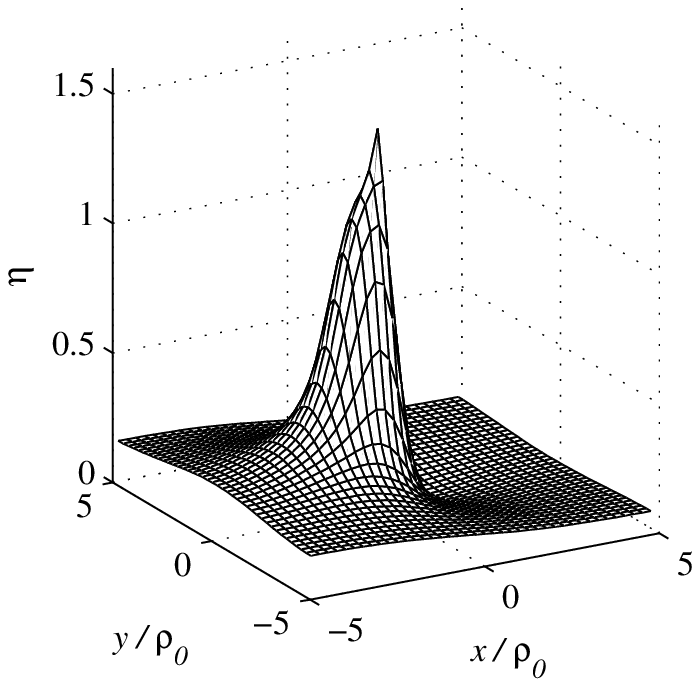}
}
\caption{Spatial distribution of  $\eta$  at various times $t=100, 150, 200, 250\Delta t$, for $R_0=-1$ (i.e.  $\Delta T<0$), and $U_1 = 0$. Only every fifth node in the mesh is used for the illustration.}
\label{fig:SurfRneg1U0Time}
\end{center}
\end{figure}

\begin{figure}[htbp]
\begin{center}
\subfigure[\hspace{1mm}]
{
\label{fig:ProfY0Rneg1U0}
\includegraphics{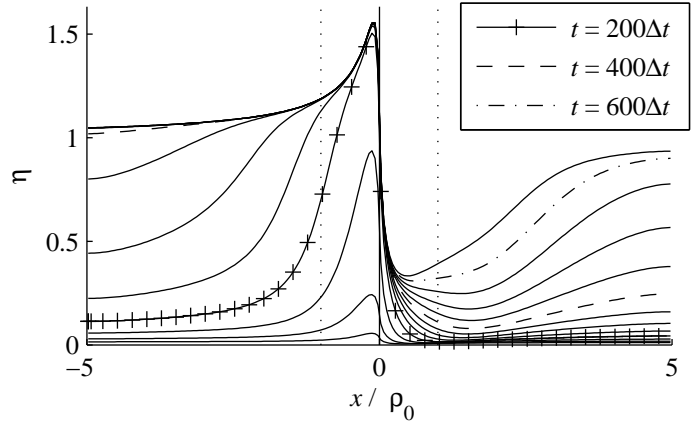}
}
\hspace{1cm}
\subfigure[\hspace{1mm}]
{
\label{fig:ContRneg1U0T7000}
\includegraphics{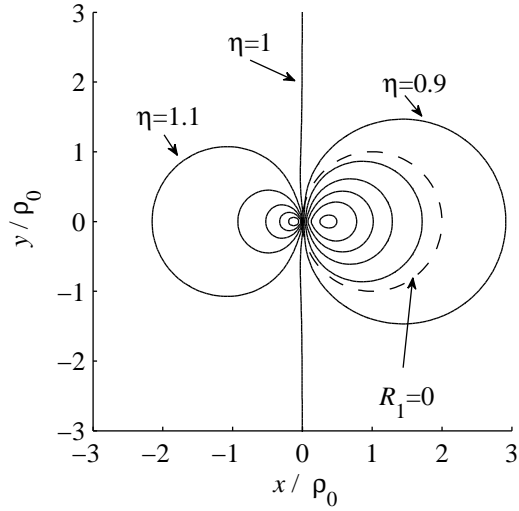}
}
\caption{$R_0=-1$ and $U_1 = 0$: (a) Evolution of $\eta(y=0$) for $t = 50 \to 650 \Delta t$. The dotted vertical lines indicate $\rho =\pm  \rho_0$; (b) Contours of $\eta$ at $t = 650 \Delta t$. At the dashed circle the sign of $R_1$ changes and is positive inside the circle.}
\label{fig:ProfCont}
\end{center}
\end{figure}

The case $R_0-1$ and $U_1=0$ corresponds to a quick decrease of temperature, i.e. quenching below the transition temperature. In Fig. \ref{fig:SurfRneg1U0Time}, the evolution of $\eta$ is illustrated by surface plots representing four different times. One can clearly see that a top is growing below the dislocation ($x < 0$) with its peak close to the dislocation. The evolution of $\eta$  is also shown in Fig \ref{fig:ProfY0Rneg1U0} as profiles of $\eta$  ($y = 0$) at different times ($t = 50 \to 650 \Delta t$ in steps of $50\Delta t$). The top first grows until the peak reaches a stable value (here $\approx 1.5$), while in the surroundings   $\eta \approx  0$ . Thereafter, the top broadens and $\eta$ increases also for $x > 0$. However, close to the dislocation, the increase is held back. This process will continue until   $\eta \approx 1$ covers the whole material, leaving a top and a valley on each side of the dislocation, respectively; see the contour plot in Fig. \ref{fig:ContRneg1U0T7000}. The dashed circle represents $R_1 = 0$, and inside it $R_1 > 0$ which, according to the Landau potential, indicates that no nucleation should take place there, cf. Fig. \ref{fig:LandauUn}. Since the magnitude of $R_1$ goes to infinity as $1/\rho$, while approaching the origin, the peak and the valley are expected to evolve. Far from the dislocation, $R_1$ goes to $R_0$, i.e. a constant negative value, and therefore nucleation is expected.

\subsection{Dislocation: $R_0 =-1$ and $U_1=-1$}
Figure 6a shows the evolution of $\eta(y = 0)$ whereupon the same trend as for $U_1 = 0$ is observed. First a top is growing near the dislocation, where an increase of $\eta$ arises from this top to finally form a plateau. However, for $R_0 =-1$ and $U_1=-1$, the evolution is faster than in the foregoing case, and both the peak and plateau values are higher, i.e.  $\eta =1.7$ and 1.3, respectively. It should be noted that the time increments are equal for all the studied cases.

\begin{figure}[htbp]
\begin{center}
\subfigure[\hspace{1mm} $U_1 = -1$]
{
\label{fig:ProfileY0Rneg1Uneg1}
\includegraphics{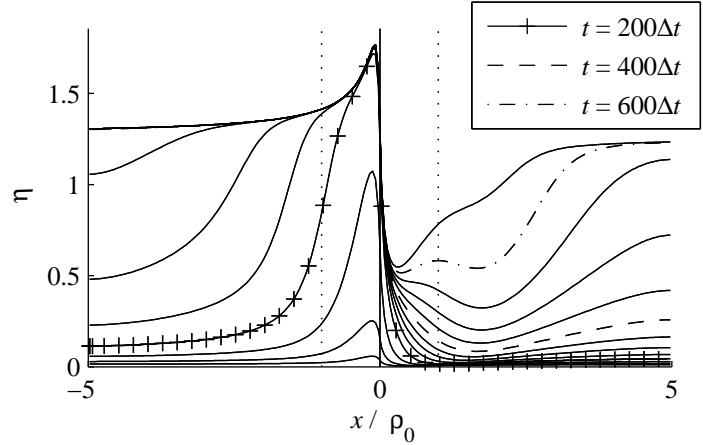}
}
\hspace{1cm}
\subfigure[\hspace{1mm} $U_1 = 1$]
{
\label{fig:ProfileY0Rneg1Upos1}
\includegraphics{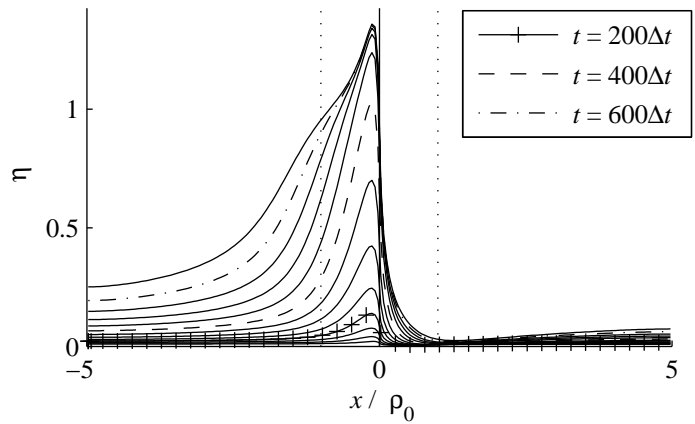}
}
\caption{Evolution of $\eta(y=0)$ for $t=50 \to 650\Delta t$, upwardly, with  $R_0=-1$. The dotted vertical lines indicate $\rho=\pm\rho_0$.}
\label{fig:ProfileY0Rneg}
\end{center}
\end{figure}
\subsection{Dislocation: $R_0 =-1$ and $U_1=1$}
In the case $R_0 =-1$ and $U_1=1$, the development of $\eta$  has the same character as the foregoing cases, with the difference that the maximum value of $\eta$  and the plateau level are smaller and the evolution is much slower. In Fig. \ref{fig:ProfileY0Rneg1Upos1}, only profiles up to $t = 650\Delta t$ are shown, however, for the evolving plateau, $\eta$ reaches around 0.8. It is observed that the plateau values for the three different cases correspond well to the real, positive roots $\eta =(1.00, 1.27, 0.78)$ that give the minima of the Landau potential for $R_1 = R_0 = -1$. These non-trivial roots are found from
\begin{equation}
\label{eqn:equilibria}
\eta_+^2  =  \frac{-U_1+\sqrt{U_1^2-4R_0}}{2}.
\end{equation}
\noindent

\subsection{Dislocation: $R_0=1$}

Now consider the situation where the temperature is above the transition temperature ($R_0 > 0$), i.e. no phase transition is expected in a non-stressed and defect free material. However, by introducing a dislocation into the system the situation alters. Figure 7 shows the evolution of $\eta$  in the case where $U_1=(-1,2,-3)$, repectively. For $U_1=-1$ a top emerges and evolves until it finds a stable shape, with $U_1=-2$ a top develops and thereafter a small plateau appears that will stop to broaden at $t \approx 2000\Delta t$, and for $U_1=-3$ the top grows and broadens until the plateau covers nearly the whole area except near the dislocation where $x < 0$. In Fig. 3a the contours where both phases have minima of the Landau potential are shown for different values of $U_1$, and the more negative is $U_1$, the larger is the circular area where the second phase is assumed to evolve. For $U_1=-1$, the diameter of the corresponding circle is approximately 1.2 and for $U_1=-2$, the diameter is 4.0. Dashed lines in the figures show the corresponding location on the $x$-axis. In the case with $U_1=-3$, the contour surrounds a small circular area below the dislocation ($x < 0$) wherein no analogous phase transformation is expected.

\begin{figure}[htbp]
\begin{center}
\subfigure[\hspace{1mm} $U_1 = -1$]
{
\label{fig:ProfileY0Rpos1Uneg1}
\includegraphics{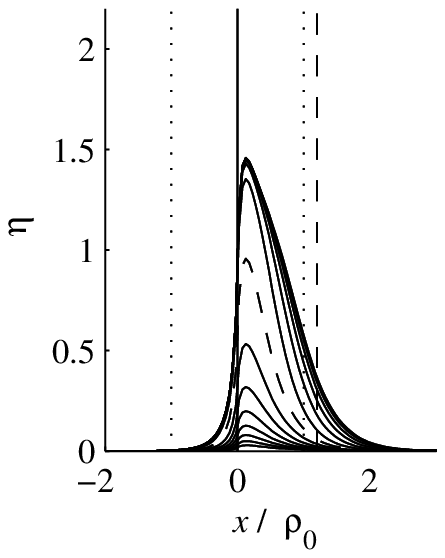}
}
\hspace{1cm}
\subfigure[\hspace{1mm} $U_1 -2$]
{
\label{fig:ProfileY0Rpos1Uneg2}
\includegraphics{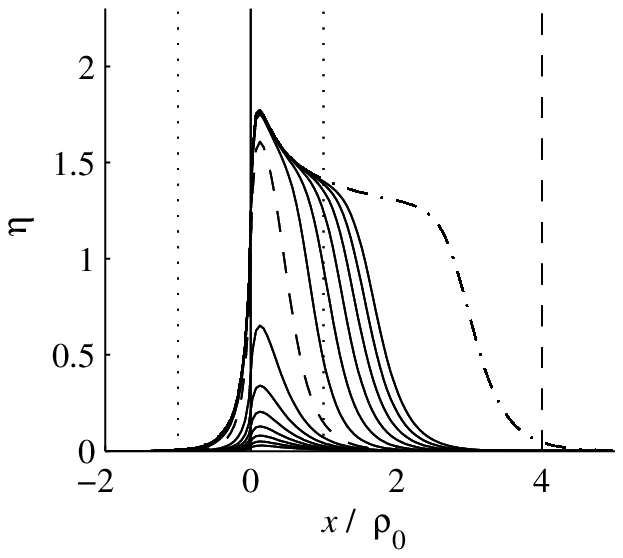}
}
\hspace{1cm}
\subfigure[\hspace{1mm} $U_1 = -3$]
{
\label{fig:ProfileY0Rpos1Uneg3}
\includegraphics{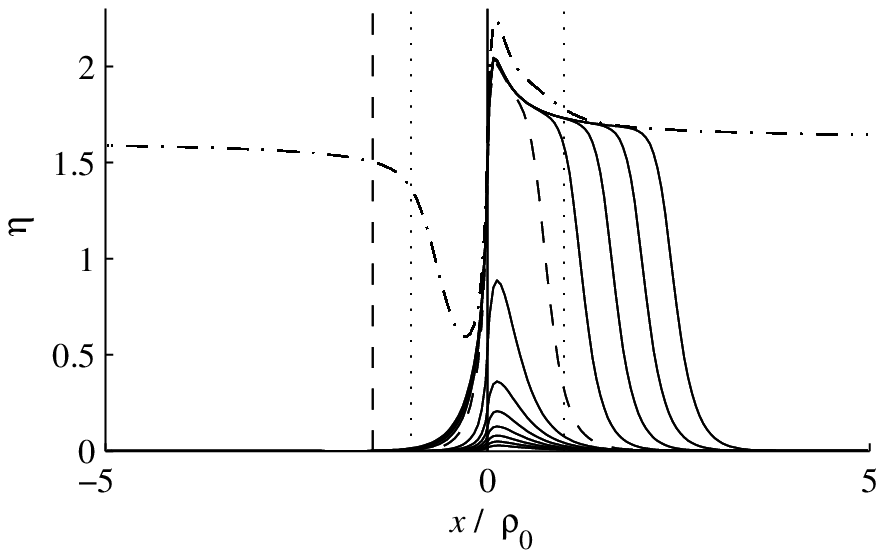}
}
\caption{Evolution of $\eta(y=0)$ for $t =50 \to 650 \Delta t$, upwardly, with $R_0 = 1$. The dashed curves represent the $\eta$ profile at $t =400 \Delta t$. In (b) and( c) profiles for $t=2000 \Delta t$ are also added (dash dot curves). The vertical dotted indicate $x =\pm \rho_0$ and the vertical dashed lines show the location with triple minima. The dashed curves represent the $\eta$ profile at $t =400 \Delta t$.}
\label{fig:ProfileY0Rpos1Uneg}
\end{center}
\end{figure}

\subsection{Crack}
Calculations have also been performed for the crack and the results are similar to that of the dislocation. In the vicinity of the crack, there is always an area where the second phase is stable regardless of the sign of $R_0$. Detailed results and analysis for the crack will be presented elsewhere.

\section{Conclusions}
\label{sec:conclude}
Numerical simulations of the spatiotemporal evolution of a second phase in the vicinity of elastic defects in crystalline solids have been performed using the Ginzburg-Landau equation for a single non-conservative structural order parameter. The computations indicate that these defects always trigger a nucleation of a second phase. In the very vicinity of the dislocation and at the crack tip a distinct top emerges and evolves. In some cases, the structural order parameter evolves into a plateau, which either finds an equilibrium shape or spreads out into the whole material. It should also be mentioned that the Landau type energy could also be used to roughly estimate the spatial evolution of the order parameter in the vicinity of defects. The study does not include the influence of concentration of species which is essential for modeling phase transition in an alloy. With a two component order parameter field the local orientation of the second phase could also be deduced. Analysis comprising the non-conserved order parameter coupled to a conserved concentration field obeying a diffusion-like equation will be presented elsewhere.

\begin{acknowledgments}
 The work was supported by the Knowledge Foundation of Sweden grant number 2008/0503.
\end{acknowledgments}

\bibliography{micro}

\end{document}